\documentclass[twocolumn,aps,prb,showpacs,superscriptaddress]{revtex4}
\usepackage[dvips]{graphicx}
\usepackage{bm}
\begin{document}
\title{First-principles calculations of the magnetism of Fe$_2$O$_2$H$_2$}
\author{Sergey Stolbov}
\email{stolbov@phys.ksu.edu} \affiliation{Department of Physics,
Kansas State University, Manhattan, KS 66506 USA}
\author{Richard A. Klemm}
\email{klemm@phys.ksu.edu} \affiliation{Department of Physics,
Kansas State University, Manhattan, KS 66506 USA}
\author{Talat S. Rahman}
\email{rahman@phys.ksu.edu} \affiliation{Department of Physics,
Kansas State University, Manhattan, KS 66506 USA}
\date{\today}
\begin{abstract}
By expanding the wave function in plane waves, we use the
pseudopotential method of  density functional theory within the
generalized gradient approximation to calculate the effective
magnetic coupling energies of the $S=5/2$ spins in the Fe2 dimer,
approximated as Fe$_2$O$_2$H$_2$.  Setting the Fe-O bond length at
the value corresponding to the minimum total energy, we find the
difference in antiferromagnetic and ferromagnetic exchange
energies as a function of  the Fe-O-Fe bond angle $\theta$.  The
effective interaction is antiferromagnetic for
63$^{\circ}<\theta<105^{\circ}$, and is ferromagnetic otherwise.
Full potential augmented plane wave calculations were also
performed at $\theta=100, 105^{\circ}$, confirming these results,
and providing information relevant to the local anisotropy of the
spin interactions.
\end{abstract}
\pacs{71.15.Mb, 61.46.+w, 75.75.+a} \vskip0pt \maketitle

Single molecule magnets (SMM's) have been under intense study
recently, due to their potential uses in magnetic storage and
quantum computing.\cite{background,sarachik,loss}  The materials
consist of insulating crystalline arrays of identical SMM's  1-3
nm in size, each containing two or more magnetic ions. Since the
magnetic ions in each SMM are surrounded by non-magnetic ligands,
the intermolecular magnetic interactions are usually negligible.
Although the most commonly studied SMM's are the high-spin  Mn12
and Fe8,\cite{background,Fe8}  such SMM's contain a variety of
ferromagnetic (FM) and antiferromagnetic (AFM) intramolecular
interactions, rendering unique fits to a variety of experiments
difficult.\cite{Fe8spin9}

 In many single molecule magnets, and in Fe8 in particular, the
magnetic core contains [LM(OR)]$_2^{+2}$ dimer ions,\cite{Fe8}
where M denotes a magnetic ion (eg. Fe$^{3+}$), L is a ligand, and
OR denotes an alkoxide ion, with R = H, CH$_3$, CH$_2$CH$_3$, etc.
In Fe8, the overall magnetic cluster is
\{[(tacn)$_6$Fe$_8$O$_2$(OH)$_{12}$]Br$_7\cdot
$H$_2$O\}[Br$\cdot$8H$_2$O], where tacn is
1,4,7-triazacyclononane.\cite{Fe8structure} Near-neighbor
Fe$^{3+}$ ions are coupled via superexchange in four distinct pair
bondings: four equivalent pairs of Fe$^{3+}$ ions are coupled
through two (OH)$^{-}$ ions with Fe-O-Fe angle
$\theta=100.7^{\circ}$, one central pair is coupled via two
O$^{2-}$ ions with Fe-O-Fe angle $\theta=104.4^{\circ}$, and each
of those Fe$^{3+}$ ions is coupled via the same O$^{2-}$ ions to
two other Fe$^{3+}$ ions, and finally, four equivalent pairings
via a single (OH)$^{-}$ ion.\cite{Fe8structure,sangregorio} Here
we focus upon the magnetic superexchange interaction between the
constituent magnetic ions that is mediated via two oxygen ions,
which is different that the usual case of superexchange via a
single oxygen ion.  The attachment of alkyl R groups to the oxygen
ions is probably of minor importance, because the oxygen orbitals
involved in the OR bond are orthogonal to those involved in the
Fe-O bonding.  The local magnetic order depends strongly upon the
Fe-O-Fe bond angle $\theta$ and the bond length, which are
determined by the local ligand environment.

In addition, Le Gall {\it et al.} synthesized and measured the
magnetization of four species of the isolated dimer Fe2,
\cite{Fe2} and electron paramagnetic resonance (EPR) experiments
were performed on one of them.\cite{Fe2expt} These dimers have the
general formula [Fe(OR)(dK)$_2$]$_2$, where  dK is a
$\beta$-diketonate ligand. In these dimers, the oxygen ion in each
alkoxide group forms a bridge between the Fe$^{3+}$ S=5/2 spins,
allowing them to interact via superexchange through both O$^{2-}$
ions, as in the [LFe(OH)]$_2^{2+}$ ions present in Fe8.  From
magnetic susceptibility measurements, fits to the isotropic
Heisenberg exchange model ${\cal H}=-J{\bm S}_1\cdot{\bm S}_2$
were made.\cite{Fe2} Note that we use the sign convention in which
a positive $J$ corresponds to ferromagnetic couplings. In
comparing these four Fe2 dimers, they found no correlation between
$J$ and the average Fe-O bridging bond distance, but a linear
correlation was found between $J$ and the Fe-O-Fe bond angle
$\theta$, with $J$ decreasing monotonically from -14.8$\pm$0.5 to
-19.0$\pm$0.6 cm$^{-1}$ as $\theta$ increased from 101.8$^{\circ}$
to 103.8$^{\circ}$, respectively. \cite{Fe2} One of these
compounds, [Fe(OMe)(dpm)$_2$]$_2$, was studied with M{\"o}ssbauer
spectroscopy.\cite{cianchi}

Le Gall {\it et al.} also compared these trends with predictions
based upon extended H{\"u}ckel calculations performed on the
simpler model system, [Fe(OH)H$_4$]$_2$,\cite{Fe2} using the
approach of Hay, Thibeault, and Hoffman that for magnetic dimers,
$J\propto\sum_i(\Delta E_i)^2$, where $\Delta E_i$ is the energy
separation between symmetric and asymmetric combinations of
coupled magnetic orbitals.\cite{HTH} Unfortunately, when they used
the oxygen orbital parameters normally expected, the $J$ values
calculated in this way increased with increasing $\theta$,
opposite to the experimental observations.\cite{Fe2}  That
approach was also not applicable for ferromagnetic exchange
couplings.\cite{HTH}

The static and dynamic properties of Fe2 dimers were studied
theoretically by Efremov and Klemm, and interesting low
temperature quantum steps in the magnetization were
predicted.\cite{ek} Recently, those authors  studied local spin
anisotropy effects, and found that the details of the low
temperature quantum magnetization steps could be complicated by
such local anisotropies.\cite{ek2}

Because of the very weak antiferromagnetic interactions
($J\approx-0.9\pm0.1$meV) between the Fe(III) spins in the two
related dimer compounds,\cite{jk,gm,gmchi,reiff,lechan}
$\mu$-oxalatotetrakis(acetylacetonato)Fe$_2$ and
[Fe(salen)Cl]$_2$, where salen is
$N,N'$-ethylenebis(salicylideneiminato), magnetization steps were
observed at low temperatures in them.\cite{shapiramu,shapiracl} In
the former case, all five of the magnetization steps were observed
using pulsed magnetic fields,\cite{shapiramu} and they were found
to be evenly spaced, suggestive of an isotropic antiferromagnetic
Heisenberg exchange interaction.\cite{ek}  In the second case, the
lower two and part of the third magnetization steps were observed
by pulsed magnetic fields at low temperature. Although the second
step was rather sharp, with a sharp $dM/dH$ peak, the first step
had a much broader linewidth, suggestive of two primary
constituents.\cite{shapiracl} Although the sample measured
contained many unoriented crystallites, the broad first peak
followed by the sharp second peak is consistent with a substantial
amount  of local spin anisotropy of the type
$-J_a\sum_{i=1}^2S_{iz}^2$, where $J_a\approx0.1J$.\cite{ek2}

Here we focus on the constituent dimers present in Fe8, which are
of the [RFeO]$_2$ type, in which each Fe$^{3+}$ ion shares
electrons with one aliphatic R group and two  bridging O$^{2-}$
ions, which have no aliphatic substitutions on them, as pictured
for R = H in Fig. 1. We model this system by replacing the R group
with H. To study the geometry effect upon the magnetic interaction
and the spin state of the dimer, we perform a first principles
calculation of the electronic and magnetic structures and of the
total energy of the [HFeO]$_2$ dimers for ferromagnetic and
antiferromagnetic states and for
$57^{\circ}\le\theta\le110^{\circ}$.  From these results, we are
able to obtain $J(\theta)$.

\begin{figure}
\includegraphics[width=0.45\textwidth]{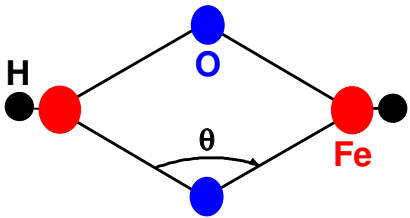}
\caption{Sketch of the [HFeO]$_2$ dimer.  The Fe-O-Fe angle is
denoted $\theta$.}
\end{figure}

The majority of the calculations presented here were performed
within the density functional theory, using the generalized
gradient approximation (GGA) for the exchange-correlation
functional,\cite{perdew} and the pseudopotential method combined
with the expansion of the wave functions in plane
waves.\cite{joannopoulos}  We assume a three-dimensional structure
of such [HFeO]$_2$ dimers, and for simplicity, pack them into a
periodic array of tetragonal unit supercells of dimension
12$\times12\times15${\AA}.  This unit supercell is sufficiently
large that the interaction between neighboring dimers is utterly
negligible.  For all atoms in the dimer, we use the ultrasoft
pseudopotentials.\cite{vanderbilt}  We set the cutoff energies for
the plane-wave expansion at 400 eV, and perform the calculaton
only for the $\Gamma$-point in the first Brillouin zone, which is
sufficient for a single molecule calculation.  To obtain the
equilibrium structure of the dimer, we apply the optimization
procedure that relaxes the system until the forces acting on each
atom converge to within 0.02 eV/{\AA}.  We obtain the values
$S_{FM}$ and $S_{AFM}$ for the local spins on the  Fe sites in
both the ferromagnetic and antiferromagnetic configurations by
integrating the local spin densities over a sphere of radius
1.1{\AA} centered about the Fe site.  For each angle $\theta$, the
total energies $E_{FM}$ and $E_{AFM}$ for the antiferromagnetic
and ferromagnetic configurations are calculated. Then, the
exchange coupling $J$ is determined from
\begin{eqnarray}
J&=&\frac{E_{AFM}-E_{FM}}{S_{AFM}S_{FM}}.
\end{eqnarray}

Ideally, if the absolute energy values $E_{FM}$ and $E_{AFM}$ were
reliable, we would calculate $J$ from
$E_{AFM}/S_{AFM}^2-E_{FM}/S_{FM}^2$.  We make the above
approximation because the absolute energy values are not nearly as
reliable  as are the much smaller energy differences.

In order to find the equilibrium structure of the dimer, we first
minimize the forces acting upon the atoms within the dimer. We
find that the minimum energy is reached for antiferromagnetic
states in which the Fe-O bond length $\ell=1.83$ {\AA}, and
Fe-O-Fe bond angle $\theta=82.4^{\circ}$. We then keep the bond
length fixed to that optimum value, and calculate $E_{FM}$,
$E_{AFM}$, $S_{FM}$, and $S_{AFM}$ for six $\theta$ values in the
range $57^{\circ}\le\theta\le 110^{\circ}$.  Our results are
presented in Table 1.

\begin{table}
\begin{tabular}{|c|c|c|c|c|}
\hline $\theta$&$S_{FM}$&$S_{AFM}$&$E_{AFM}-E_{FM}$&$J$\cr
(deg)&&&(meV)&(meV)\cr \hline 57&2.81&2.95&260&31\cr
70&2.79&3.37&-456&-48\cr 80&2.88&3.41&-645&-66\cr
90&2.55&3.23&-415&-50\cr 100&3.63&3.65&-170&-13\cr
110&3.34&3.39&64&6\cr \hline
\end{tabular}
\caption{Magnetic characteristics of the [HFeO]$_2$ dimer
calculated using the pseudopotential method for different Fe-O-Fe
angles $\theta$.}
\end{table}

We find that for $\theta>105^{\circ}$ and for $\theta<63^{\circ}$,
the dimer prefers to be in a ferromagnetic ground state, whereas
for $63^{\circ}<\theta<105^{\circ}$, an antiferromagnetic coupling
is preferred.  The change in magnetic order is accompanied by
strong changes in the spin density on the Fe ions, as well as in
the spin density on the O ions that control the superexchange
interactions, which data are not presented.  For the Fe dimer
components in Fe8, the [LFe(OH)]$_2^{2+}$ effective dimers have
$\theta=100.7^{\circ}$, which our calculations indicate is likely
to be antiferromagnetic, contrary to the experiments.  The central
effective [LFeO]$_2^{2+}$ dimer has $\theta=104.4^{\circ}$, which
is borderline between ferromagnetic and antiferromagnetic in our
calculations.  Experimentally, it is ambiguous, because the ground
state configuration of those spins are also affected by
superexchange via a single oxygen in an OH$^{-}$ ion, which is
likely to be strongly antiferromagnetic.

In Fig. 2, we have plotted our $J(\theta)$ data, along with the
exerimental data of Le Gall {\it et al.} for the four Fe2
dimers.\cite{Fe2}  We note that the experimental data points are
close to the guide to the eye between our calculated angles
$\theta=100,100^{\circ}$, so that our pseudopotential calculation
is close to predicting the experimentally observed exchange
energies of these four compounds.

\begin{figure}
\includegraphics[width=0.45\textwidth]{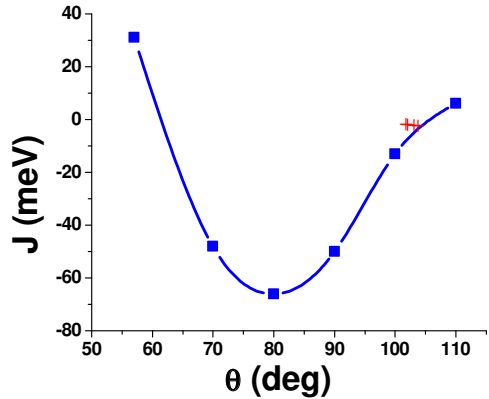}
\caption{Plot of the exchange energy $J$ in meV versus the Fe-O-Fe
angle $\theta$ in degrees.  Shown are our results from Table 1
(solid blue square), a guide to the eye (solid blue curve), and
the data for Fe2 dimers (red crosses).\cite{Fe2}}
\end{figure}

We remark that Pederson, Kortus, and Khanna have performed a much
more accurate set of calculations for Fe8, and obtain the
experimentally found  S=10 ground state, with the central
[LFeO]$_2^{2+}$ dimer being antiferromagnetically coupled to the
other Fe spins.\cite{pederson}  They also found global anisotropy
parameters in rather close agreement with experiment as deduced
from magnetization experiments and from oscillations in the
quantum tunnel splitting.\cite{carciuffo,tunnel}

We note that although the dimer we have studied is rather
different from the Fe2 dimers that exhibit superexchange via the
oxygen ions in alkoxides, to the extent that the alkoxides can be
approximated by the O$^{2-}$ ions we studied, the $\theta$ values
observed in the four Fe2 dimers varied from 101$^{\circ}$ to
104$^{\circ}$, which are in the antiferromagnetic
regime.\cite{Fe2} As in the extended H{\"u}ckel calculations, it
appears that the general trend we calculate of increasing $J$ with
increasing $\theta$ is opposite to the experiments on the four Fe2
dimers.  However, we do not have enough data points between
100$^{\circ}$ and 105$^{\circ}$ to determine if the general trend
is accurately followed.  More important, there appear to be
serious discrepancies with the interesting case of
[Fe(salen)Cl]$_2$, with $\theta\approx90^{\circ}$ and a very weak,
antiferromagnetic exchange
interaction.\cite{gm,gmchi,reiff,lechan,shapiracl}

In addition, our pseudopotential calculation does not contain any
specific information regarding the relative spin densities in the
various occupied Fe orbitals.  Hence, we are unable to determine
the local spin anisotropy parameters relevant for more detailed
studies of Fe2 dimers.\cite{ek2} In order to obtain useful
information regarding the local spin anisotropies, another
calculational procedure, such as a full potential augmented plane
wave calculation, would be necessary.  In [Fe(salen)Cl]$_2$, the
unusual behavior of the first magnetization step is strongly
suggestive of a substantial local spin anisotropy interaction, as
noted above.\cite{shapiracl,ek2}

As a preliminary check upon the validity of the pseudopotential
method and also as a first-principles investigation of the
possibility of local spin anisotropy, we have made full potential
augmented plane wave (FLAPW) calculations of the spin densities
and energies of the [HFeO]$_2$ dimer at $\theta=100, 105^{\circ}$.
In Tables II and III, we presented our results obtained using the
FLAPW method for the partial charges in the various Fe(III)
$d$-orbitals for both up ($\uparrow$) and down ($\downarrow$) spin
configurations in the FM and AFM configurations, on Fe1 and Fe2
sites, for $\theta=100, 105^{\circ}$, respectively.  In Table IV,
the energy differences between the FM and AFM cases for these
angles are also presented.

\begin{table}
\begin{tabular}{|c|c|c|c|c|c|c|}\hline
Fe $d$&FM&FM & AFM&AFM&AFM&AFM\cr orbital&Fe1,2 $\uparrow$&Fe1,2
$\downarrow$&Fe1 $\uparrow$&Fe1 $\downarrow$&Fe2 $\uparrow$&Fe2
$\downarrow$\cr \hline
$d_{z^2-r^2}$&0.8454&0.2541&0.8451&0.2194&0.2192&0.8451\cr
$d_{x^2-y^2}$&0.8822&0.4177&0.8875&0.0875&0.0885&0.8875\cr
$d_{xy}$&0.8953&0.0549&0.8930&0.1321&0.1312&0.8931\cr
$d_{xz}$&0.7494&0.2585&0.8777&0.3247&0.3247&0.8776\cr
$d_{yz}$&0.8910&0.0562&0.8923&0.0921&0.0918&0.8923\cr \hline
\end{tabular}\caption{FLAPW results for the partial charges within the Fe
$d$-orbitals for $\theta=100^{\circ}$.  The FM and AFM cases for
up and down spins within the  Fe1 and Fe2 spheres are presented.}
\end{table}

\begin{table}
\begin{tabular}{|c|c|c|c|c|c|c|}\hline
Fe $d$&FM&FM & AFM &AFM&AFM &AFM\cr orbital&Fe1,2 $\uparrow$&Fe1,2
$\downarrow$&Fe1 $\uparrow$&Fe1 $\downarrow$&Fe2 $\uparrow$&Fe2
$\downarrow$\cr \hline
$d_{z^2-r^2}$&0.8407&0.2511&0.8431&0.2173&0.2172&0.8430\cr
$d_{x^2-y^2}$&0.8830&0.4215&0.8911&0.0670&0.0669&0.8911\cr
$d_{xy}$&0.8949&0.0613&0.8924&0.1490&0.1496&0.8924\cr
$d_{xz}$&0.7555&0.2490&0.8849&0.3102&0.3102&0.8849\cr
$d_{yz}$&0.8914&0.0533&0.8920&0.0936&0.0937&0.8920\cr \hline
\end{tabular}\caption{FLAPW results for the partial charges within the Fe
$d$-orbitals for $\theta=105^{\circ}$.  The FM and AFM cases for
 up and down spins within the   Fe1 and Fe2 spheres are presented.}
\end{table}

\begin{table}
\begin{tabular}{|c|c|}\hline
$\theta$ (deg)& $E_{AFM}-E_{FM}$ (eV)\cr\hline 100&-0.69134\cr
105&1.0594\cr \hline\end{tabular}\caption{FLAPW calculations of
the energy differences between AFM and FM states for
$\theta=100,105^{\circ}$} \end{table}

\begin{table}
\begin{tabular}{|c|c|c|c|c|c|c|}\hline
$\theta$&FM &FM &AFM1 &AFM1 &AFM2 &AFM2\cr
(deg)&$Q_{\uparrow}$&$Q_{\downarrow}$&$Q_{\uparrow}$&$Q_{\downarrow}$&$Q_{\uparrow}$&
$Q_{\downarrow}$\cr\hline
 100&8.49303&5.16213&8.63626&4.98165&4.98117&8.63650\cr
105&8.49901&5.15439&8.64531&4.96278&4.96329&8.64520\cr
\hline\end{tabular} \caption{Total charge densities on the Fe 1
and 2 sites for up and down spins, respectively, for the FM and
AFM configurations, at the two angles studied using the FLAPW
technique.} \end{table}

 In comparing our FLAPW results from Table IV with
those obtained from the pseudopotential method in Table I, we see
that they both predict a crossover from AFM behavior for
$\theta\le100^{\circ}$ to FM behavior at larger angles.  Although
we did not perform the pseudopotential calculation at
$\theta=105^{\circ}$, our FLAPW calculation at 105$^{\circ}$
predicts FM behavior, so that the predicted crossover at
$\theta_c$ is for $100^{\circ}<\theta_c<105^{\circ}$.  In
addition, it is evident that  the largest differences in the spin
densities at within the $d_{xy}$ and $d_{yz}$ orbitals for the FM
case for both $\theta$ values, whereas comparable (for
$\theta=100^{\circ}$) or slightly larger (for
$\theta=105^{\circ}$) differences in the spin densities occur
within the $d_{x^2-y^2}$ orbitals for the AFM configurations,
respectively.  We expect that the bonding orbitals will be the
$d_{xz}$ and $d_{z^2-r^2}$ orbitals, for which the magnitudes of
the differences in the spin densities were found to be nearly
equivalent for both the AFM and FM configurations, for both angles
studied. Thus, the spin density differences  within the bonding
orbitals, that participate in the exchange interactions, depend
very weakly upon the Fe-O-Fe bond angle $\theta$.  The greatest
$\theta$ dependence is for spin density differences in the
non-bonding orbitals, and the orbital that shows the greatest
distinction between FM and AFM behavior is the $d_{x^2-y^2}$
orbital.  Thus, it appears that the $d_{x^2-y^2}$-orbital governs
the local spin anisotropy in Fe2 dimers.

In order to estimate the strength of the effective Heisenberg
coupling, if we take the differences in the total spin densities
from Table V, we obtain $J=-57$ meV and $J=86$ meV for
$\theta=100,105^{\circ}$, respectively.  Taking the spins to have
the value 4 in each case, we get $J=-43$ meV and $J=66$ meV,
respectively.  These numbers are larger in magnitude that the
values obtained using the pseudopotential technique.

We remark that the most interesting Fe2 dimer to date is
[Fe(salen)Cl]$_2$, which has Cl ions in place of the H ions
pictured in Fig. 1.  In addition, there are organic ligands
attached to the O ions, as in all of the Fe2 dimers presently
known.\cite{Fe2,gm,shapiracl}  We have not yet studied this very
interesting case, but plan to do so soon.  The Cl$^{-}$ ions could
polarize the local spins in the Fe $d$-orbitals, and it would be
interesting to see if the $d_{z^2-r^2}$ orbitals would be the most
affected.

 This work was
supported in part by the NSF through contract NER-0304665.

\end{document}